\documentclass[12pt]{article}
\usepackage[utf8]{inputenc}

\usepackage{amsfonts,amsfonts,amsmath,amssymb,bm} 
\usepackage[parfill]{parskip} 
\usepackage{color}
\usepackage{graphicx}
\usepackage{natbib}
\usepackage[english]{babel} 
\usepackage[margin=3.cm]{geometry}
\usepackage{enumitem}
\usepackage{hyperref}
\usepackage{subcaption} 
\usepackage{float} 
\usepackage{authblk} 
\usepackage[stretch=10]{microtype} 
\usepackage[section]{placeins} 

\setcitestyle{numbers,square,comma}

\numberwithin{equation}{section}

\makeatletter

\newcommand{\beq}{\begin{equation}}
\newcommand{\eeq}{\end{equation}}

\providecommand{\keywords}[1]
{
  \small	
  \textbf{\textit{Keywords---}} #1
}

\begin{document}

\title{Space-time statistics of a linear dynamical energy cascade model}

\author[1]{Gabriel B. Apolin\'ario\thanks{gabriel.brito\_apolinario@ens-lyon.fr}}
\author[1]{Laurent Chevillard}
\date{}

\affil[1]{Univ Lyon, ENS de Lyon, Univ Claude Bernard, CNRS, Laboratoire de Physique, 46 all\'ee d'Italie F-69342 Lyon, France}

\maketitle

\begin{abstract}
A linear dynamical model for the development of the turbulent energy cascade was introduced in Apolin\'ario \emph{et al} (J. Stat. Phys. \textbf{186}, 15 (2022)). This partial differential equation, randomly stirred by a forcing term which is smooth in space and delta-correlated in time, was shown to converge at infinite time towards a state of finite variance, without the aid of viscosity. Furthermore, the spatial profile of its solution gets rough, with the same regularity as a fractional Gaussian field. We here focus on  the temporal behavior and derive explicit asymptotic predictions for the correlation function in time of this solution and observe that their regularity is not influenced by the spatial regularity of the problem, only by the correlation in time of the stirring contribution. We also show that the correlation in time of the solution depends on the position, contrary to its correlation in space at fixed times. We then investigate the influence of a forcing which is correlated in time on the spatial and time statistics of this equation. In this situation, while for small correlation times the homogeneous spatial statistics of the white-in-time case are recovered, for large correlation times homogeneity is broken, and a concentration around the origin of the system is observed in the velocity profiles. In other words, this fractional velocity field is a representation in one-dimension, through a linear dynamical model, of the self-similar velocity fields proposed by Kolmogorov in 1941, but only at fixed times, for a delta-correlated forcing, in which case the spatial statistics is homogeneous and rough, as expected of a turbulent velocity field. The regularity in time of turbulence, however, is not captured by this model.
\end{abstract}


\section{Introduction}

Since the important work of Taylor and Richardson in the first half of the XX century, a direct energy cascade has been invoked to interpret the rough nature of velocity fluctuations that are observed in Navier-Stokes turbulence in three dimensions \cite{frisch1995}. This picture has been further supplemented by the statistical description provided by Kolmogorov in 1941 \cite{kolmogorov1941}, in which fractional stochastic processes play a paradigmatic role in modeling the rough behavior of incompressible velocity fields. The energy cascade can then be seen as the dynamical process through which energy is transported from the large scale smooth external force to the small scales of the flow, where it generates rough random fields. Furthermore, it is conjectured that this rough stationary state is reached even in the absence of viscous dissipation \cite{onsager1949,constantin1994,duchon2000,eyink2006}.

Consider that $u_i(x,t)$, $i \in \{1,2,3\}$, $x \in \mathbb{R}^3$ is a divergence-free velocity field, solution of the incompressible Navier-Stokes equations. The statistical behavior of turbulence in \cite{kolmogorov1941}, based on two propositions regarding symmetries and self-similarity of these solutions, describes a fluid of kinematic viscosity $\nu$, stirred at a large scale $L$ by an external force, and determines a power-law scaling for the statistical moments of increments of the velocity field (also called structure functions of order $q$),
\beq \label{eq:K41}
\lim_{\nu \to 0} \mathbb{E} (\delta_{\ell} u_i)^q \underset{\ell \to 0}{\sim} c_q \left(\frac{\ell}{L}\right)^{q H} \ ,
\eeq
where the velocity increments are defined as $\delta_{\ell} u_i(t,x) = u_i(t,x+\ell) - u_i(t,x)$, and the constants $c_q$ do not depend on the exact setup of the flow, but only on the large scale $L$ and on the mean energy dissipation rate. The exponent $H$ is equal to $1/3$, as predicted by dimensional analysis and verified by numerical and experimental measurements \cite{frisch1995,iyer2017,debue2018,dubrulle2019}.
It should be remarked that the characterization of Eq.~\eqref{eq:K41} has been later supplemented by the intermittency hypothesis, in which the structure function exponents do not grow linearly with the order of the statistical moments. This means that the underlying velocity field is not exactly self-similar, but rather a multifractal, displaying values of the exponent $H$ which fluctuate in space and time around the value of $1/3$ \cite{frisch1995}.

In this setting, the fractional Gaussian field of exponent $H=1/3$ can be seen as the simplest model of a turbulent velocity field, with statistics given by Eq.~\eqref{eq:K41}.
These mathematical objects, introduced by Kolmogorov (in the framework we henceforth refer to as K41) and named by Mandelbrot and Van Ness \cite{mandelbrot1968,robert2008,lodhia2016}, are uniquely characterized by the H\"older exponent $H$, which describes their local regularity, coinciding with the behavior of Brownian motion for $H=1/2$, while being rougher than that for $H < 1/2$ and smoother for $H > 1/2$.
Nevertheless, being Gaussian and of mean zero, all of the odd order increment moments of fractional Gaussian fields vanish, in contrast with the phenomenology of turbulence, where strong skewness has been observed in the statistical distribution of velocity increments, and where an important exact result on the third order structure function has been obtained, given by $\mathbb{E}(\delta_{\ell} u_i)^3 = -\frac45 \varepsilon \ell$, with $\varepsilon$ being the mean energy dissipation rate \cite{frisch1995,kolmogorov1941dissip}. This discrepancy has lead to several proposals for synthetic random fields with a local regularity prescribed either by Eq.~\eqref{eq:K41} or by its intermittent extensions \cite{schmitt2001,pereira2016,skewed2019,friedrich2021}.

Nevertheless, these models are static, and do not address the fundamental dynamics of the formation of small scales from the random advection of large scale flows, established by the external forcing. Some success in the heuristics of turbulent statistics from a dynamical description has been obtained in the framework of shell models \cite{gledzer1973,ohkitani1989,biferale2003}, which are systems of nonlinear differential equations for the variables $u_k$, called shells, inspired by the Navier-Stokes equations in Fourier space. Each shell has a characteristic wavelength $k$ and interacts only with its closest neighbors in a discretized, often logarithmic, $k$-space, as a way to model the local transport of energy in Fourier space observed in turbulence.
With constraints such as energy and helicity conservation, and a with large scale forcing (i.e. a forcing concentrated around shells of wavelength $k_0$, where $k_0$ is small), a scaling behavior very similar to that of Navier-Stokes is observed, albeit only in a numerical context.

The present work addresses the modeling of self-similar random velocity fields in a dynamical framework, building upon the results of \cite{apolinario2021}, in which a nonlinear partial differential equation (PDE) for a complex one-dimensional velocity field was introduced. The stationary solution of this PDE, when subject to a random force on the large scales, is multifractal and fractional, as controlled respectively by the intermittency parameter $\gamma$ and by the H\"older parameter $H$. Multifractal behavior is included as a nonlinear interaction governed by $\gamma$, which is a small number. If this parameter is taken to be zero, a linear PDE is obtained and its stationary state is a fractional Gaussian field of parameter $H$.
In this linear PDE, small scales are built dynamically by an homogeneous operator of degree 0 \cite{attractors2020,dyatlov2019,deverdiere2020}, which has been identified in the phenomenon of concentration of internal waves around geometrical attractors in stably stratified flows \cite{maas1997,rieutord1997,scolan2013,brouzet2016},
and whose discretized version has been used to describe a linear cascade of energy in the context of shell models \cite{mattingly2007}.

This work concentrates on the linear PDE introduced in Ref. \cite{apolinario2021}, which describes a complex velocity field $u_{H,\nu}(t,x)$, defined on $x \in \mathbb{R}$ and characterized by a positive kinematic viscosity $\nu$ and a H\"older exponent $H \in ]0,1[$. This velocity field is governed by
\beq \label{eq:frac}
\partial_t u_{H,\nu}(t,x) = P_H \mathcal{L} P_H^{-1} u_{H,\nu}(t,x) + \nu \partial_x^2 u_{H,\nu}(t,x) + f(t,x) \ , \eeq
where temporal and spatial derivatives are denoted respectively by $\partial_t$ and $\partial_x$. A viscous dissipation term, proportional to the viscosity $\nu$ and an external large scale force $f$ are present.
The operator $\mathcal{L}$ on the right-hand side of Eq.~\eqref{eq:frac} is identified with the homogeneous operator of degree 0 described in \cite{attractors2020,dyatlov2019,deverdiere2020} and its action on a general velocity field $u(t,x)$ is defined as
\begin{equation}
\mathcal{L} u(t, x) \equiv 2 i \pi c x u(t, x) \ ,
\end{equation}
where $c$ is a constant.
This term is responsible for the transport of energy from the large scale $L$ to the small scales, where viscous dissipation acts strongly in regularizing the velocity field.
It can be seen that the dynamics induced by $\mathcal{L}$ is a linear transport in Fourier space, towards increasing wavelengths in the case of $c>0$. In this paper, we always consider $c$ to be positive, but the case of $c<0$ is equivalent, with the energy cascade flowing to large negative wavelengths. Ultimately, the operator $P_H$, responsible for the local regularity of order $H$ of the stationary state solution, is defined by
\beq \label{eq:PHop} P_H u(t,x) = \int_{\mathbb{R}} e^{2\pi i k x} \frac{1}{|k|_{1/L}^{H+1/2}} \widehat{u}(t,k) \, dk \ , \eeq
where $|k|_{1/L}$ is a regularization of the norm $|k|$ on small wavelengths, whose exact description is not crucial, and $\widehat{u}$ represents a Fourier transform of $u$, with the chosen definition
\begin{equation}
\mathcal{F}[u](t, k) \equiv \widehat{u}(t, k)=\int_{\mathbb{R}} \, e^{-2 i \pi k x} u(t, x) \, dx \ .
\end{equation}

The combined action of $\mathcal{L}$ in generating a delta-correlated velocity field and of $P_H$ in building power-law correlations from this delta-correlated field lead to the formation of a stationary state which resembles a fractional Gaussian field, as explained in \cite{apolinario2021}.

The forcing contribution of Eq.~\eqref{eq:frac} is taken to be a complex Gaussian random field of mean zero and covariance
\beq \label{eq:fcov} \mathbb{E}[f(t,x)f^*(t',y) ] = \chi(t-t') C_f(x-y) \ , \eeq
where $*$ represents complex conjugation. Moreover, the real and imaginary components of $f$ are independent, meaning that $\mathbb{E}\left[f(t, x) f(t', y)\right]=0$. $\chi$ and $C_f$ are even functions of their arguments, representing the correlations in time and in space of the external force, respectively. These functions have finite characteristic scales, a length $L$ for $C_f$ and a time $T$ for $\chi$, beyond which they rapidly vanish. The scale $L$ is furthermore identified in the turbulence phenomenology as the integral length scale, at which energy is injected in the flow.

The case of white in time forcing (that is, $\chi(t)$ is a Dirac delta function) has been considered in \cite{apolinario2021} and is revisited in this work, where we investigate the properties in time and in space of the solution $u_{H,\nu}$. Additionally, we examine the case of a finite correlation time $T$ and establish a connection with the previous results in the limit $T \to 0$.
We choose the initial condition $u_{H,\nu}(0,x) = 0$, and, for the correlation functions, we specify
$\chi(t) = e^{-|t|/T} / 2 T$
and
$C_f(x) = e^{-x^2/2 L^2}$ as particular cases which have been applied to the numerical simulations.
Focusing on the linear case, Eq.~\eqref{eq:frac}, for which analytical solutions can be found, allows us to fully understand the asymptotic temporal and spatial behavior of these velocity fields.

This paper is organized as follows. In Sec. 2, we discuss the statistical properties in space and in time of the solution of Eq.~\eqref{eq:frac} driven by a delta-correlated external force, with analytical results and numerical simulations. In Sec. 3, these statistical properties are revisited for a force of finite correlation time, with numerical simulations and analytical results. Concluding remarks and future perspectives are drawn in Sec. 4.

\section{Dynamical fractional Gaussian fields}
\label{sec:fgf}

In this section we discuss the statistical properties of the stationary solution of Eq.~\eqref{eq:frac} under delta-correlated forcing in time. We begin with recalling the spatial statistics that were first analyzed in \cite{apolinario2021}, and we then introduce statistical properties in time as well. This is also an opportunity to present the details of our numerical simulations and directly compare them with the analytical results.

\subsection{Statistical properties for a delta-correlated in time forcing term}

\subsubsection{Spatial statistics}

As discussed in \cite{apolinario2021}, the solution of Eq.~\eqref{eq:frac}, $u_{H,\nu}$, reaches a statistically stationary state of finite variance even for vanishing viscosity, and spatial slices of this stationary solution resemble fractional Gaussian fields. At zero viscosity, the system only reaches this exactly self-similar description at infinite time, since the external energy injected at scale $L$ by the forcing term is linearly transported in $k$-space to higher frequencies, and an infinite amount of time is necessary for arbitrarily small wavelengths to be manifest. In the presence of a finite viscosity, though, wavelengths higher than some typical dissipative wavelength are strongly suppressed, and the system reaches its stationary state in a finite time. Nevertheless, since both the solutions at finite or zero viscosity are statistically stationary, their statistical features are very similar, differing only in the dissipative range of wavelengths, where viscosity regularizes the rough local behavior of the solution. Therefore, we discuss the solution of the inviscid fractional equation
\beq \label{eq:frac-0}
\partial_t u_{H,0}(t,x) = P_H \mathcal{L} P_H^{-1} u_{H,0}
(t,x) + f(t,x) \ ,\eeq
with initial condition $u_{H,0}(t=0,x)=0$ and $f$ a large scale Gaussian forcing, smooth in space, exactly as in Eq.~\eqref{eq:fcov}. We first consider the delta-correlated case, $\chi(t) = \delta(t)$.
This equation admits an analytical solution in its Fourier space formulation,
\beq \label{eq:frac-solution}
\widehat u_{H,0}(t,k) = |k|_{1/L}^{-H-\frac12} \int_0^t |k-cs|_{1/L}^{H+\frac12} \widehat f(t-s, k-cs) ds \ ,
\eeq
for any forcing scheme. 
The small wavelength regularization might be defined as $|k|_{1/L} = (k^2 + L^{-2})^{1/2}$, a choice which is important in the numerical simulations.
The solution in Eq.~\eqref{eq:frac-solution} is Gaussian since it is defined as a linear operation on the forcing $f$, which is itself chosen Gaussian. Hence, to characterize the solution from a statistical point of view, it suffices to investigate second order observables, namely the power spectrum and the structure function, which elucidate the nature of the energy cascade and of the local regularity of the velocity field.

In its stationary state, obtained as the limit of $t \to \infty$, Eq.~\eqref{eq:frac-solution} is a rough velocity field, correlated over the large scale $L$. Its local regularity can be described by the power-law decay of the power spectrum density, defined as 
\beq \widehat{C}_{H,0}(t,k) = \mathcal{F} \, \mathbb{E}[u_{H,0}(t,x+\ell) u_{H,0}^*(t,x)](t,k) \ , \eeq
where the Fourier transform is performed over the separation variable $\ell$, since the correlation function in this equation is homogeneous, as can be seen from the analytical solution.
The exact expression for the power spectrum of $u_{H,0}$ is
\beq \label{eq:delta-exact-spec} \widehat{C}_{H,0}(t,k) = |k|_{1/L}^{-(2H+1)} \int_0^t |k-cs|_{1/L}^{2H+1} \widehat{C}_f(k-cs) ds \ , \eeq
which in the limit of large wavelengths, in the stationary state, becomes 
\beq \label{eq:frac-spec} \lim_{t \to \infty} \widehat{C}_{H,0}(t,k) \underset{k \to \infty}{\sim} b_H \, |k|^{-(2H+1)} \eeq
with a constant given by
\beq b_H = \frac{1}{c} \int_{\mathbb{R}} |s|_{1/L}^{2H+1} \widehat{C}_f(s) ds \ , \eeq
an expression which can be explicitly evaluated in terms of special functions in the case of a Gaussian correlation function $\widehat{C}_f$,  and can also be easily numerically integrated in a general setting.
Eq.~\eqref{eq:frac-spec} illustrates the direct energy cascade, in which higher wavelengths are progressively excited, linearly in time.
As previously stated, the variance of the solution $u_{H,0}$ is finite, even with a vanishing viscosity. It is given by
\beq \label{eq:delta-var} \mathbb{E}|u_{H,0}|^2 = \lim_{t \to \infty} \int_{\mathbb{R}} \widehat{C}_{H,0}(t,k) \, dk \ , \eeq
and its finiteness is a consequence of the integrability of the power spectrum.

The local regularity of the solution is also measured by the small scale asymptotic statistics of the increments $\delta_{\ell} u_{H,0}(t,x)$, defined as
\beq \delta_{\ell} u_{H,0}(t,x) = u_{H,0}(t,x+\ell) - u_{H,0}(t,x) \ . \eeq
The increment variance, or second order structure function, is asymptotically given by the power-law relation
\beq \label{eq:frac-sf2-space} \lim_{t \to \infty} \mathbb{E}|\delta_{\ell} u_{H,0}|^2 \underset{\ell \to 0^+}{\sim} c_H \, \ell^{2H} \ , \eeq
where the constant is
\beq \label{eq:frac-sf2-spaceMultConstant}
c_H = \frac{(2\pi)^{2H+1}}{2 c \sin(\pi H) \Gamma(1+2H)} \int_{\mathbb{R}} |s|_{1/L}^{2H+1} \widehat{C}_f(s) \, ds \eeq
and $\Gamma$ represents the Gamma function. Higher order expectations of the increments, $\mathbb{E}(\delta_{\ell} u_{H,0})^q$, can be obtained directly from simple combinatorial arguments, since the increments are Gaussian as well. Furthermore, $q$-order expectation values display an exponent linear in $H$, and the expectation value themselves vanish for all odd $q$, as a consequence of the vanishing mean of the velocity field.

This representation of a turbulent velocity field is thus compatible with the portrait of Eq.~\eqref{eq:K41}. With the H\"older parameter chosen as $H=1/3$, the power-law decay in Eq.~\eqref{eq:frac-spec} has an exponent $-5/3$ and the increment variance has an exponent $2/3$, two values which are widely regarded as universal in the statistical description of turbulence, and which are given by the K41 framework. Nonetheless, we remark that the multifractal description of turbulence induces a small deviation to the value $1/3$, which requires the introduction of an intermittency parameter. The nonlinear equation addressed in \cite{apolinario2021}, where the intermittency parameter governs the intensity of the nonlinearity, induces non-Gaussian statistics on its solution, which capture both the multifractal behavior and the nontrivial nature of the odd order structure functions of real turbulent velocity fields.

\subsubsection{Temporal statistics}

We now turn to the statistical properties in time of the inviscid solution. We observe that the $P_H$ operator does not significantly influence the temporal behavior of $u_{H,0}$, and that its regularity in time is solely governed by the shape of the~ forcing correlation in time. The regularity in time is again investigated through increments $\delta_{\tau} u_{H,0}(t,x)$, analogously defined as
\beq \delta_{\tau} u_{H,0}(t,x) = u_{H,0}(t+\tau,x) - u_{H,0}(t,x) \ . \eeq
It is possible to write the following expression for the temporal second-order structure function,
\beq \label{eq:delta-sf2-exact}
\lim_{t \to\infty} \mathbb{E}|\delta_{\tau} u_{H,0}(t,x)|^2
= \lim_{t \to\infty} 2 \mathbb{E}|u_{H,0}(t,x)|^2 - 2 \mathcal{R} \, \mathbb{E}[u_{H,0}(t+\tau,x)u_{H,0}^*(t,x)] \ , \eeq
in terms of the variance and the real part of the two-point correlation-function in time, equal to
\beq \begin{split} \label{eq:delta-corrf}
\lim_{t \to\infty} \mathbb{E}[u_{H,0}(t+\tau,x)u_{H,0}^*(t,x)]
&= \frac{1}{c} \int_{k\in\mathbb{R}} e^{2\pi i c\tau x} |k|_{1/L}^{-H-1/2}|k+c\tau|_{1/L}^{-H-1/2} \\&\times \int_{s=-\infty}^k |s|_{1/L}^{2H+1} \widehat{C}_f(s) \, ds dk \ . \end{split} \eeq
This equation shows an explicit oscillatory behavior with the position $x$, which is inherited by the structure function. To see the behavior of the temporal second-order structure function at small scales $\tau\to 0$, we perform a Taylor expansion of the integrand and get
\beq \label{eq:delta-sf2-time-off}
\lim_{t \to\infty} \mathbb{E}|\delta_{\tau} u_{H,0}(t,x)|^2 = d_{1,H} \, |\tau| + d_{2,H}(x) \, \tau^2 + \mathcal{O}(\tau^3) \ , \eeq
where the constants are given by
\beq \label{eq:delta-sf2-consts} \begin{split}
&d_{1,H} = (2H+1) \int_{k\in\mathbb{R}} k |k|_{1/L}^{-2H-3} \int_{s=-\infty}^k |s|_{1/L}^{2H+1} \widehat{C}_f(s) \, ds dk \ \mbox{, and} \\
&d_{2,H}(x) = 4 c \pi^2 x^2 \int_{k\in\mathbb{R}} |k|_{1/L}^{-2H-1} \int_{s=-\infty}^k |s|_{1/L}^{2H+1} \widehat{C}_f(s) \, ds dk \\
&+ \frac{c}{4} (2H+1) \int_{k\in\mathbb{R}} \left[ 2/L^2 - (2H+3)k^2 \right] |k|_{1/L}^{-2H-5} \int_{s=-\infty}^k |s|_{1/L}^{2H+1} \widehat{C}_f(s) \, ds dk
\ . \end{split} \eeq
Eq.~\eqref{eq:delta-sf2-time-off} says that the spatio-temporal field $u_{H,0}(t,x)$ has the same local regularity in time as a Brownian motion, as given by the linear term in the expansion. Remark that, contrary to its spatial behavior, where it has been pinpointed a regularity similar to a fractional Brownian motion of parameter $H$ (Eq.~\ref{eq:frac-sf2-space}), the regularity in time is independent of $H$.
It can be observed that the integrals in Eqs.~\eqref{eq:delta-sf2-time-off} and \eqref{eq:delta-sf2-consts} are finite given that the function $\widehat{C}_f$ decays fast enough. A Gaussian decay for instance, which is employed in the numerical simulations, is sufficient.
Furthermore, $d_{1,H}$ is positive, since it can be written as an integral over positive terms only, and the first contribution of $d_{2,H}$ is already expressed as an integral of positive terms. The second term of $d_{2,H}$, which does not depend on $x$, can be shown to be positive for the specific regularization we employ through an integration by parts, even though its exact value is hard to obtain. We reinforce that the $d_{1,H}$ term is independent of $x$ and of $c$, which play a role only in the second order contribution.

An important quantity in the temporal dynamics of $u_{H,0}$ is its integral correlation time, which is usually defined as a time integral of the two-point correlation function in time (Eq.~\ref{eq:delta-corrf}), normalized by the variance. Nevertheless, it can be observed that Eq.~\eqref{eq:delta-corrf} is not integrable at $\tau \to \infty$ for $H < 1/2$. For this reason, in order to obtain a finite value of the integral correlation time, we propose an estimation based on the Taylor expansion of the second-order structure function in time. In the limit of large time intervals, Eq.~\eqref{eq:delta-sf2-time-off} is no longer valid and the temporal second-order structure function evolves to the homogeneous value of twice the variance of $u_{H,0}$, as can be seen in the exact expression, Eq.~\eqref{eq:delta-sf2-exact}. We then define the position-dependent correlation time $T_H(x)$ as the value in which the quadratic expansion equals the asymptotic value at large times. The positive solution of this quadratic equation is
\beq \label{eq:delta-int-time-off} T_H(x) = \frac{-d_{1,H} + \sqrt{ d_{1,H}^2 + 8 d_{2,H}(x) \mathbb{E}|u_{H,0}|^2}}{2 d_{2,H}(x)} \ , \eeq
which elicits that the asymptotic behavior at large $x$ of $T_H(x)$ is a decay inversely proportional to $x$.
Since all terms in the above equation are finite and positive, $T_H(x)$ is positive for all $x$.
Furthermore, a numerical evaluation of $T_H(0)$ confirms the expectation from dimensional analysis that this quantity is approximately inversely proportional to the integral length scale $L$.

These results show that the statistical properties in space at fixed time of the fractional linear cascade model, Eq.~\eqref{eq:frac}, reproduce well the fractional Gaussian field. On the other hand, statistical properties in time are inhomogeneous, in contrast with the phenomenology of stochastic Navier-Stokes turbulence, in which homogeneous statistics is observed both for time and spatial increments. In turbulent velocity fields, the Eulerian structure function in time (that is, increments in time measured at some fixed point in space) show the same local regularity as their spatial equivalent, with a power-law structure function approximately given by (for the second order) $\mathbb{E}(\delta_{\tau} u)^2 \propto \tau^{2/3}$ in the limit of vanishing viscosity, for a small time interval $\tau$. This effect, which has been measured in direct numerical simulations \cite{chevillard2005, gorbunova2021}, can be phenomenologically explained with the sweeping effect \cite{tennekes1972}, the random advection of the small scales by the large scale motion of the flow, and reproduced with spatio-temporal random fields \cite{chaves2003,reneuve2020}.

We briefly mention that the aforementioned properties hold as well for the viscous fractional equation, Eq.~\eqref{eq:frac},
and can be derived from its analytical solution \cite{apolinario2021}.
With a finite viscosity, the solution develops a finite inertial range where the behavior described by the spatial statistics of Eqs.~\eqref{eq:frac-spec} and \eqref{eq:frac-sf2-space} is valid, but which is interrupted by the dissipative length introduced by the viscosity, of the order of $(\nu/c)^{1/3}$. With the action of dissipation, the system furthermore reaches its stationary state in a finite time of the order of $1/(c^{4/3} \nu^{1/3})$. One other consequence of the dissipative term is regularizing the solution at small spatial scales, creating a dissipative range in which the second order structure function in space is proportional to $\ell^2$, as expected for a smooth function. This regularization in space however is not transmitted to the time dependence, which remains with the same roughness as a Brownian motion.

\subsection{Numerical results} \label{sec:delta-numerical}

Numerical simulations of Eq.~\eqref{eq:frac} have been performed with a pseudo-spectral method in space and a first order predictor-corrector scheme for the time evolution, as described in Ref.~\cite{apolinario2021},  where the same numerical method has been applied to the study of the equivalent nonlinear equation. An important difference in the method is that a spatial truncation of the force is applied in \cite{apolinario2021}, through a smooth bump function which vanishes at the boundaries of the domain, $\pm L_{\mathrm{tot}}/2$. This truncation is fundamental in the nonlinear setting, since the simulation domain is finite and thus the operator $\mathcal{L}$ is discontinuous at these boundaries. The interaction of the nonlinearities with this discontinuous operator leads to a numerical divergence which is avoided by the truncation. The simulations performed in this work do not employ such truncation, since the linear equation is much more well behaved numerically. An advantage of this approach, without the truncation, is that the velocity field is homogeneous, as expected from the theory, and entire time slices of it can be used for the evaluation of statistical observables, easily generating large statistical ensembles. The effects of the discontinuous operator are small, but still noticeable, and will be discussed in the next paragraphs.

The chosen parameters for our simulations are the following: The spatial dimension, with a total length of $L_{\mathrm{tot}} = 1.0$ is composed of $N=2^{12}$ points, with a spacing of $\Delta x = N/L_{\mathrm{tot}}$, which leads to a maximum wavelength of $k_{\mathrm{max}} = N/ 2L_{\mathrm{tot}}$. For the external force, the functional form $C_f(x) = \exp(-x^2/2 L^2)$ was chosen, with three different values of the correlation length, $L=( L_{\mathrm{tot}}/8, L_{\mathrm{tot}}/16, L_{\mathrm{tot}}/32)$. 
Three values were chosen for the H\"older exponent as well, $H = (1/3, 1/2, 2/3)$, where $H=1/3$ is adopted for most figures because of its prominence in the statistics of turbulence. $H=1/2$ represents a standard Brownian motion and $H=2/3$ is a random field which is more regular than Brownian motion. 
In order to show the action of the dissipation in regularizing small scales, the following values of viscosity are chosen: $\nu = (2 \times 10^{-7}, 3 \times 10^{-8}, 2 \times 10^{-9}, 2 \times 10^{-10})$. 
We set the initial condition to $u_{H,\nu}(0,x) = 0$ and the simulation first runs for a transient time large enough for the system to reach its statistically stationary state. This transient time is estimated as $T_{\mathrm{stat}} = 2^{12}$, since the transport rate can be set to $c=1.0$ without loss of generality. Since the $\mathcal{L}$ operator induces a linear transport in Fourier space, this transient time is linearly proportional to the dissipative wavelength of the system and the chosen value is enough for it to reach a statistically stationary state. After this transient regime, the simulation runs for more $T_{\mathrm{tot}} = 2^{12}$ units of time, and during this stage we store the data for later analysis as several velocity profiles in space at different instants in time and as profiles in time at different positions. The time steps were taken as $\Delta t = \Delta x$, as in \cite{apolinario2021}, a value which is seen to produce well-resolved velocity fields, since the PDE we investigate is linear with a very small viscosity. For larger values of the viscosity, a choice of $\Delta t \propto (\Delta x)^2$ would have been necessary, as expected from the stability analysis of the heat equation (the dissipative term) \cite{canuto2007}.

\begin{figure}[t]
\centering
\includegraphics[width=\textwidth]{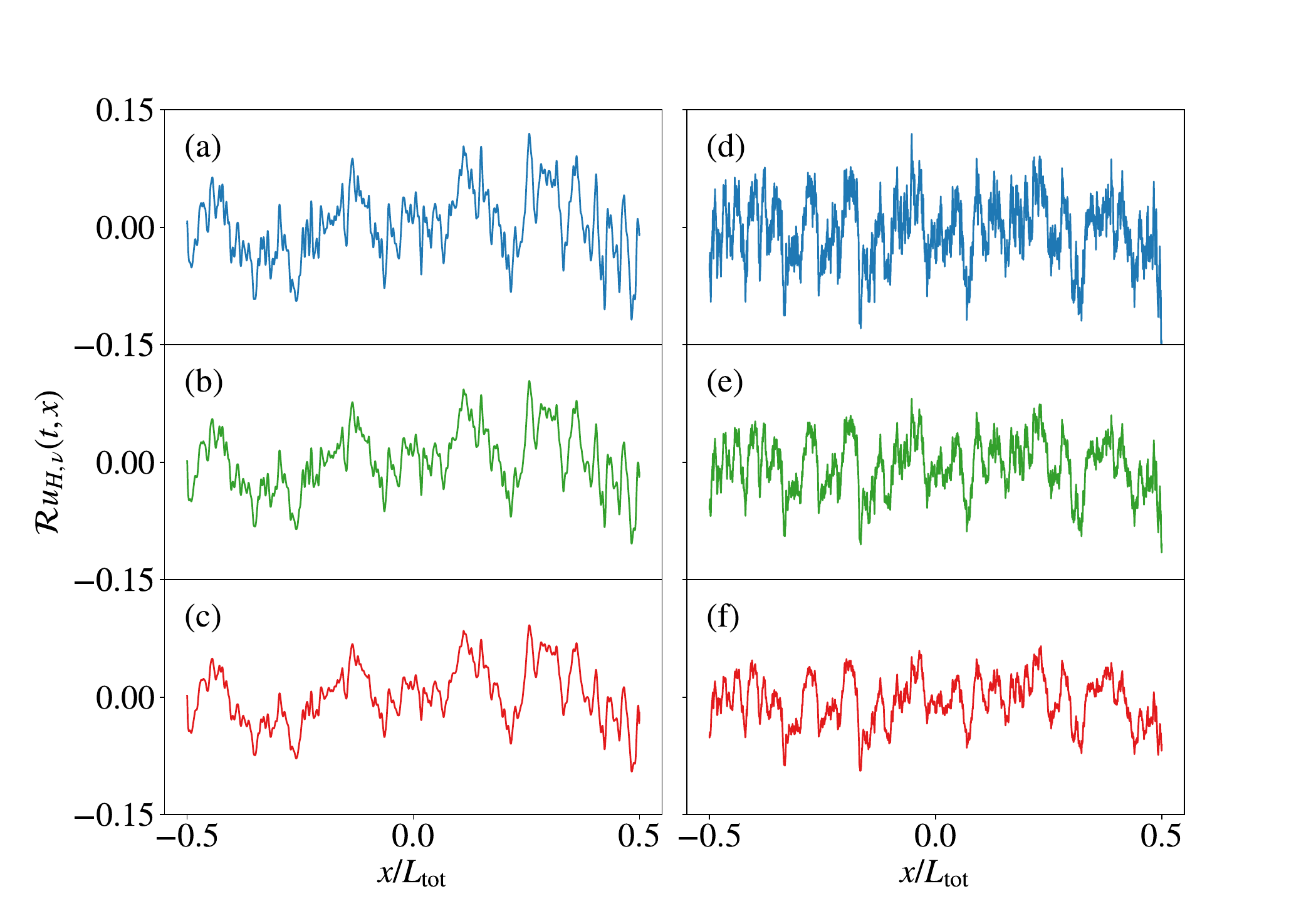}
\caption{Spatial profile of the real part of the numerical solution of Eq.~\eqref{eq:frac} at a fixed time $t$ in the statistically stationary regime for different values of the parameter $H$, for $N=2^{12}$, $L=L_{\mathrm{tot}}/8$ and $c=1.0$. In (a,b,c), the viscosity is $\nu=1 \times 10^{-8}$ and in (d,e,f) it is $\nu=1 \times 10^{-11}$.
The different H\"older exponents are $H=1/3$ (blue, in a and d), $H=1/2$ (green, in b and e) and $H=2/3$ (red, in c and f).
For each value of viscosity, the same random space-time realization of the forcing was used for the three values of $H$.}
\label{fig:delta-space-profile}
\end{figure}

In Fig.~\ref{fig:delta-space-profile}, the real part of the spatial profile of the velocity field $u_{H,\nu}$ at a fixed time is shown for two different values of the viscosity, $\nu = 2 \times 10^{-7}$ (a) and  $\nu = 2 \times 10^{-10}$ (b), for all three values of $H$, shown with different colors. The same random instance of the external forcing was applied to all three velocity fields for each fixed viscosity, which explains the fact that all three velocity fields at either (a) or (b) look very similar, since Eq.~\eqref{eq:frac} is linear. Nevertheless, the curves display different roughness based on their values of the parameter $H$, with the blue curve ($H=1/3$) being the one with highest fluctuations, while the red one ($H=2/3$) is the smoothest. The velocity fields look homogeneous over the whole system, since the forcing is not truncated, and a discontinuity at the boundary can be noticed, due to the discontinuity in the $\mathcal{L}$ operator. Furthermore, the effect of the large difference in viscosity between (a) and (b) is noticeable in the fact that Fig.~\ref{fig:delta-space-profile}(b) develops more roughness and smaller scales than at (a). Yet, both velocity fields share the same large scale correlation length $L=L_{\mathrm{tot}}/8$, noticeable in the large scale behavior of all curves. The chosen value for $L$ as a few fractions of $L_{\mathrm{tot}}$ allows for the easier visualization of the spatial correlations in $u_{H,\nu}$, but this value must be small enough such that the correlation function $C_f(x)$ is well resolved and sufficiently small at the boundaries of the system, otherwise an extra numerical discontinuity would be introduced.
We have only shown the real part of the velocity fields since the imaginary part looks similar.

\begin{figure}[htb]
\centering
\includegraphics[width=.86\textwidth]{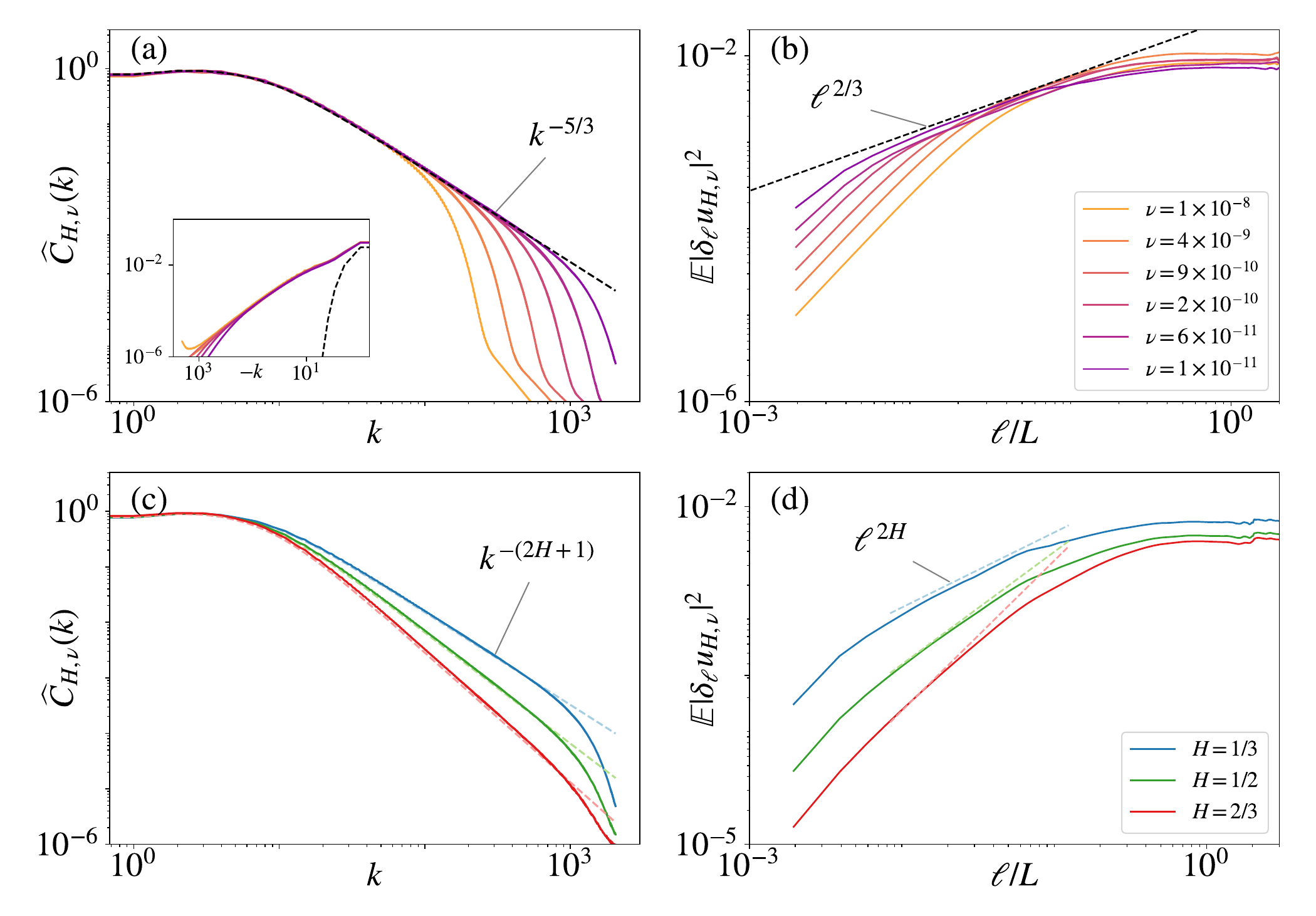}
\caption{Numerical estimations of the power spectrum (a,c) and second order structure function in space (b,d) of the solution $u_{H,\nu}$ with the same parameters as in Fig.~\ref{fig:delta-space-profile}.
In (a) and (b), only one value of the H\"older exponent, $H=1/3$, is presented, with different colors representing different viscosities, following the legend. In (c) and (d), on the other hand, three different values of $H$ are displayed, with a fixed viscosity $\nu=1 \times 10^{-11}$. The dashed black lines in (a) and (b) show the asymptotic analytical expressions, for the inviscid equation, of the power spectrum (Eq.~\ref{eq:delta-exact-spec}) in (a) and of the increment variance (Eq.~\ref{eq:frac-sf2-space}) in (b), together with their corresponding power-law dependence indicated. The inset shows the negative part of the power spectrum and its analytical result. The same analytical expressions are shown in (c) and (d), in dashed lines and colored according to the respective $H$ value.}
\label{fig:delta-space-stats}
\end{figure}

The second order statistical properties in space of the velocity field $u_{H,\nu}(t,x)$ are depicted in Fig.~\ref{fig:delta-space-stats}, namely the power spectrum (a,c) and the second order structure function in space (b,d). The estimation of the power spectrum is obtained from the periodogram, as the square norm of the discrete Fourier transform of the numerical solution divided by $N$, averaged over several independent instants in time, obtained from a single long realization of Eq.~\eqref{eq:frac}. In (a), the periodogram is shown for different values of the viscosity in different colors, as indicated in the legend, along with the analytical expression for the power spectrum of the inviscid velocity field (Eq.~\ref{eq:delta-exact-spec}), in a black dashed curve.
It can be observed that the numerical curves exactly follow the analytical result up to some maximum wavelength where the numerical power spectrum rapidly decays. Furthermore, this maximum wavelength grows as the viscosity is reduced, and for the smallest value of viscosity a large inertial range with a power-law decay of $k^{-(2H+1)}$ can be observed. We mention that, since the viscous fractional equation also possesses an explicit analytical solution, we can also plot its analytical power spectrum, and observe that it matches the numerical curve not only on the energy containing and inertial range, but also on the dissipative range (data not shown), up to a new maximum wavelength where boundary effects in the simulations can be noticed. These boundary effects can be seen in Fig.~\ref{fig:delta-space-stats}(a) as a change of slope in the dissipative range. This is a consequence of the discontinuity of the operator $\mathcal{L}$ due to the finite size of the system in the numerical simulations, but we remark that it only differs from the analytical results in the far dissipative range. The inset of Fig.~\ref{fig:delta-space-stats}(a) shows the negative part of the spectrum, which is not symmetric in $k$, a consequence of the complex nature of the velocity field $u_{H,\nu}$. This part of the spectrum is dominated solely by the forcing correlation function and decays fast, as can be seen in the numerical or in the analytical result. Nevertheless, the effect of the discontinuity can be more clearly noticed as the accumulation of energy at negative frequencies, as seen in the inset. This deviation from the analytical results leads to a numerical variance which is slightly bigger than the analytical result, since the variance is an integral of the power spectrum (Eq.~\ref{eq:delta-var}), this difference is however numerically quite small. Additionally, the inverse cascade due to the discontinuity of $\mathcal{L}$ has been analogously observed in Ref.~\cite{dyatlov2019}, where this operator is replaced by a periodic version, as $2 i \pi c x \to i L_{\mathrm{tot}} \sin(2 \pi c x / L_{\mathrm{tot}})$, but the regions of negative slope in the sine produce a strong accumulation of energy at the boundaries. For these reasons, the original operator is directly employed in our simulations without periodization.

In Fig.~\ref{fig:delta-space-stats}(b) the second order structure function in space is shown, as a function of the separation distance $\ell$, rescaled by the integral length scale $L$, in log-log scale. 
To estimate this quantity, we average the square norm of the increment, $\delta_{\ell} u_{H,\nu}(t,x) = u_{H,\nu}(t,x+\ell) - u_{H,\nu}(t,x)$, over independent instants in time and in space as well. This velocity field is homogeneous, hence a large number of samples can be obtained.
An inertial range with an $\ell^{2H}$ power law behavior, which agrees with the analytical predictions at vanishing viscosity (black dashed line) can be noticed, although it is smaller than the inertial range for the power spectrum. At small separation, the dissipative range manifests as an $\ell^{2}$ behavior in the structure function, signaling a smooth velocity field. It can be observed as well that the transition from the inertial range to the dissipative range happens at smaller scales for the smallest values of viscosity. At large separations, all curves converge to twice the value of the variance of the velocity field, where the small difference in the variance of each curve is due to dissipation.
In order to verify the dependence of the spatial statistics on $H$, Fig.~\ref{fig:delta-space-stats}(c,d) shows the power spectrum and second order structure function of $u_{H,\nu}$ for a single value of the viscosity ($\nu=1 \times 10^{-11}$) and three different values of the H\"older parameter. Different power-law exponents can be observed in the inertial range for each value of $H$, and they closely follow the analytical predictions of Eqs.~\eqref{eq:delta-exact-spec} and \eqref{eq:frac-sf2-space}, indicated in dashed colored lines.

\begin{figure}[t]
\centering
\includegraphics[width=\textwidth]{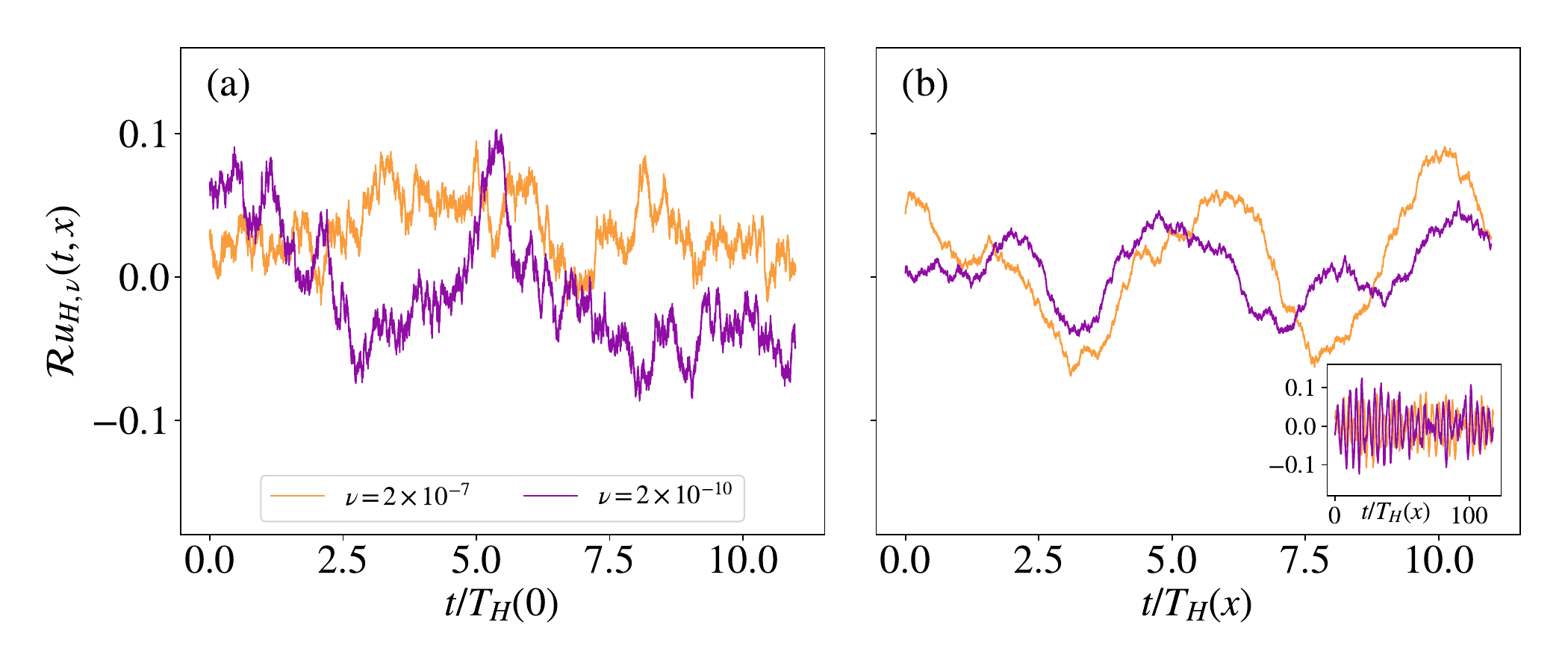}
\caption{Profile in time of the real part of the numerical solution at its stationary state at two different positions, with the same parameters as in the previous figures and $H=1/3$. In (a), the evolution in time is shown at the origin of the system, and in (b), at $x=0.1 L_{\mathrm{tot}}$. Each plot shows two values of viscosity, a large one and a small one. The plots have been rescaled in terms of the position-dependent integral correlation time of the velocity field, as given by the estimation in Eq.~\eqref{eq:delta-int-time-off}, and show equivalent intervals after this rescaling.
The inset of (b) shows the profile in time at $x=0.1 L_{\mathrm{tot}}$ but for the same total time as in (a), in the units of the simulation.}
\label{fig:delta-time-profile}
\end{figure}

We now turn to a discussion of the statistical properties in time of $u_{H,\nu}$ when driven by a delta-correlated force.
In Fig.~\ref{fig:delta-time-profile}, the profile in time of the real part of the velocity field is shown, at two different positions, $x=0$ (a) and $x = L_{\mathrm{tot}}/10$ (b), with two different values of viscosity in each plot, as indicated in the legend.
The regularity in time of this random process is that of a Brownian motion, independently of the spatial H\"older parameter, as shown by Eq.~\eqref{eq:delta-sf2-time-off}. For this reason, only $H=1/3$ is shown, with a similar behavior observed in other cases. We also observe that viscous dissipation does not affect the small time scales, since both curves at each plot are roughly indistinguishable in a statistical sense, showing the same correlation time, amplitude of fluctuations and level of roughness. The horizontal axes in (a) and (b) have been rescaled in terms of the local integral correlation time, according to Eq.~\eqref{eq:delta-int-time-off}, and it can be seen that the large scale patterns are similar in both figures.
Since $T_H(x)$ at (b) is a few times smaller than $T_H(0)$, we also show, in the inset of (b), the profile in time of the numerical solution for the same total time as in (a), in units of the simulation. In this inset, many multiples of the integral time behavior can be noticed. This reinforces the relevance of a position-dependent integral time scale, since the large-scale behavior in (a) and the inset of (b) show strong qualitative differences.

\begin{figure}[t]
\centering
\includegraphics[width=.86\textwidth]{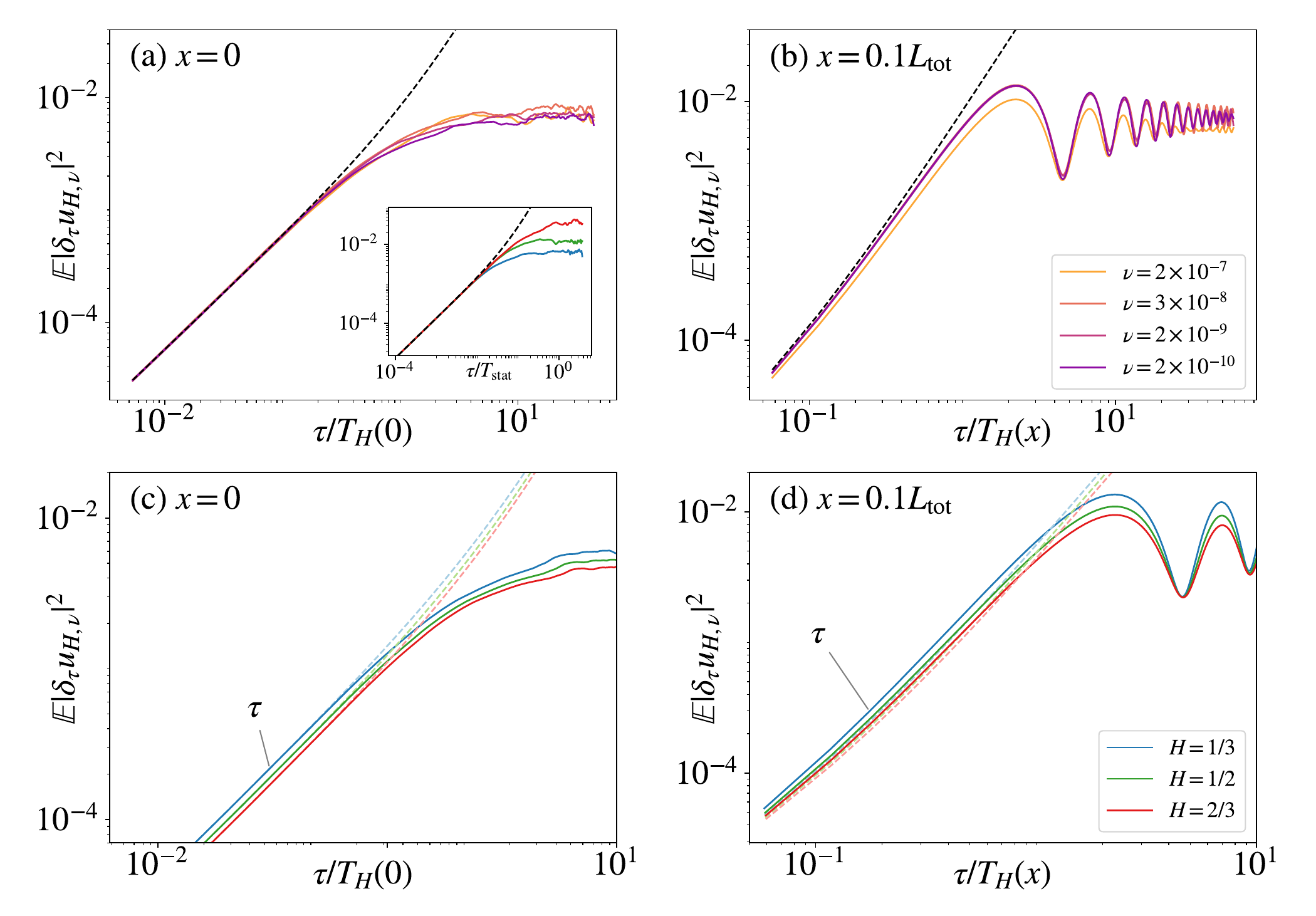}
\caption{Estimations of the second order structure function in time of $u_{H,\nu}$ at its stationary state. In (a) and (b), $H=1/3$ and different values of viscosity $\nu$ are shown, as shown in the legend, while in (c) and (d), viscosity is fixed at $\nu=2 \times 10^{-10} $ and three different values of $H$ are shown. In (a) and (c), the structure function is measured at the origin of the system and, in (b) and (d), at $x=0.1 L_{\mathrm{tot}}$. Dashed lines indicate asymptotic analytical results, given by Eq.~\eqref{eq:delta-sf2-time-off}, and each plot is rescaled by its respective position-dependent integral correlation time $T_H(x)$. In the inset of (a), the second order structure function in time, at the origin, for $\nu=2 \times 10^{-10}$ and different forcing correlation lengths $L$, is shown, with the following values: $L_{\mathrm{tot}}/8$ (blue), $L_{\mathrm{tot}}/16$ (green) and $L_{\mathrm{tot}}/32)$ (red). The inset has not been rescaled by $T_H(0)$, since it depends on $L$, but by $T_{\mathrm{stat}}$, which depends only on the viscosity.}
\label{fig:delta-time-stats}
\end{figure}

The nonhomogeneities of the velocity profile in time are inherited by the temporal second order structure function, which is shown in  Fig.~\ref{fig:delta-time-stats} for two different positions, the same as in the previous figure. In Fig.~\ref{fig:delta-time-stats}(a), the analytical prediction at small time separations given by Eq.~\eqref{eq:delta-sf2-time-off} is plotted in a black dashed curve, with the coefficients $d_{1,H}$ and $d_{2,H}$ evaluated numerically. It can be seen that, at small times, the analytical curve matches all numerical observations with different values of viscosity, confirming the claim that the local regularity in time is determined only by the correlations in time of the force and not by viscosity. The analytical curve, furthermore, has a slope of one for small times, corresponding to the local regularity of a Brownian motion. 	The concavity of the black curve, nonetheless, is positive, since it was obtained from a quadratic approximation, and to capture the negative concavity of the numerical curves a higher order expansion in $\tau$ would be required. The inset of (a) shows the same observable for the smallest viscosity and three different values of $L$, along with the analytical prediction which does not depend on $\nu$ or $L$. The correlation length influences the value of the variance, leading thus to different correlation times, as given by Eq.~\eqref{eq:delta-int-time-off}. Moreover, the reasoning behind the approximation taken to obtain Eq.~\eqref{eq:delta-int-time-off} is visible in the inset, where we observe a rapid transition from the linear behavior to the large separation constant behavior.
In Fig.~\ref{fig:delta-time-stats}(b), the same analytical prediction is seen (Eq.~\ref{eq:delta-sf2-time-off}), calculated at $x=0.1 L_{\mathrm{tot}}$. The numerical curves are seen to agree on the positive concavity of the analytical approximation, for small values of $\tau$. Just as in Fig.~\ref{fig:delta-time-profile}, a different rescaling is used at (a) and (b), corresponding to each local integral time scale, calculated numerically.
In the inset, the chosen rescaling is $T_{\mathrm{stat}}$, which does not depend on $L$, but only on the viscosity, unlike the correlation time $T_H(0)$ which depends on $L$ but not on the viscosity.
In contrast to the spatial statistics, the power-law behavior of the time structure functions does not depend on the value of $H$, as shown in Fig.~\ref{fig:delta-time-stats}(c,d). They display an $H$-dependent prefactor but the same behavior proportional to $\tau$, as can be seen in the numerical (continuous curves) and asymptotic analytical (dashed colored) results.

\section{Correlated in time forcing} \label{sec:corrf}

The natural limit to contrast the results of a delta-correlated forcing would be that of a frozen force, a random force correlated in space on a length $L$ but constant in time. Nevertheless, it can be shown that the variance of $u_{H,0}$ at the origin, that is, $\lim_{t \to \infty} \mathbb{E}|u_{H,0}(t,x=0)|^2$, diverges under a frozen force if the H\"older parameter is smaller than $1/2$. This means the dynamical evolution of Eq.~\eqref{eq:frac} is not able to regularize a constant forcing scheme, producing a state of infinite variance at the origin which is unphysical. As we have said, the range $H < 1/2$ is of great interest due to its connection with the phenomenology of turbulence, and for this reason we concentrate in this section on a forcing scheme with finite correlation time $T$. We choose the correlation function in time as
\beq \label{eq:time-corrf} \chi(t) = \frac{1}{2T} e^{-|t|/T} \ , \eeq
such that the results of the previous section can be reobtained in the limit of $T \to 0$, in which the above correlation function is reduced to a Dirac delta.
In the next section, we revisit analytical and numerical results of Eq.~\eqref{eq:frac}, now driven by an exponentially correlated in time forcing. 
As will be discussed, the statistical homogeneity of the spatial observables is broken, a strong qualitative difference in comparison to the delta-correlated forcing case. We discuss the time statistics as well, which were already not homogeneous under a white-in-time forcing.

\subsection{Numerical results}

Under a correlated forcing in time, a numerical scheme for deterministic differential equations must be employed to obtain $u_{H,\nu}$, since the external force is now continuous in time and space. For this reason, we generate the forcing with a numerical scheme for stochastic differential equations, and for the time evolution of the system, given the forcing, we use a fourth-order Runge-Kutta scheme. 
The stochastic scheme for the forcing is the same as in the previous section, a first-order predictor-corrector algorithm \cite{kloeden2013numerical}.
A forcing with the specific correlation profile of Eq.~\eqref{eq:fcov} is provided by an Ornstein-Uhlenbeck process, defined by the solution of the following dynamics:
\beq \label{eq:ornstein-uhlenbeck}
df(t,x) = - \frac{1}{T} f(t,x)dt + \frac{1}{T} d\eta(t,x) \ . \eeq
To completely specify the random force $d\eta$, we furthermore say it is of zero mean, smooth in space and delta-correlated in time, with independent real and imaginary parts. These conditions can be stated precisely in terms of the integration against a smooth test function $h(t)$ as
\beq \mathbb{E}\int h(t) d\eta(t,x) = 0 , \eeq
\beq \mathbb{E}\int h(t_1) h(t_2) d\eta(t_1,x) d\eta(t_2,y) = 0 \ \mathrm{and} \eeq
\beq \mathbb{E}\int h(t_1) h(t_2) d\eta(t_1,x) d\eta^*(t_2,y) = C_f(x-y) \int h^2(t) dt , \eeq
where $C_f(x)$ is chosen the same as in Eq.~\eqref{eq:fcov}.
The drift and diffusion terms entering in Eq. \ref{eq:ornstein-uhlenbeck} have been chosen such that the correlation function of the solution, $f$, in its statistically stationary state, is
\beq  \mathbb{E}[f(t +\tau,x) f^*(t,y)] = \frac{1}{2 T} e^{-|\tau|/T} C_f(x-y) \ , \eeq
as stated before.
The correlations in space of the force are obtained from a circulant embedding algorithm, that is, the convolution of independent complex variables (of mean zero and variance $\sqrt{dx/2}$ for their real and imaginary parts) with a Gaussian kernel, which produces a random series with Gaussian autocorrelation function.
To begin the simulations, the force must be at its stationary state, a process that takes a time of the order of $T$. After this first transient phase, the force is then used in the numerical solution of Eq.~\eqref{eq:frac} in the same manner a deterministic external contribution would be. Despite the different algorithm, at this stage the simulation occurs as described in Sec. \ref{sec:delta-numerical}: There is a second transient phase determined by the linear transport in Fourier space, which takes a time of the order of $T_{\mathrm{stat}} \propto \nu^{-1/3}$, and after that a stationary phase, from which we collect random series in space and in time.
The parameters chosen for these numerical simulations are similar to the ones already used: $N=2^{12}$, $L_{\mathrm{tot}}=1.0$, $L = L{\mathrm{tot}}/8$, $c=1.0$, $\nu = 2 \times 10^{-10}$ and $\Delta t = \Delta x$. The same three values of the H\"older parameter were used, $H=1/3$, $H=1/2$ and $H=2/3$, and for the forcing correlation time the following values were chosen: $T = (0.1, 0.2, 0.5, 2.0, 5.0, 10.0)$.

\begin{figure}[t]
\centering
\includegraphics[width=\textwidth]{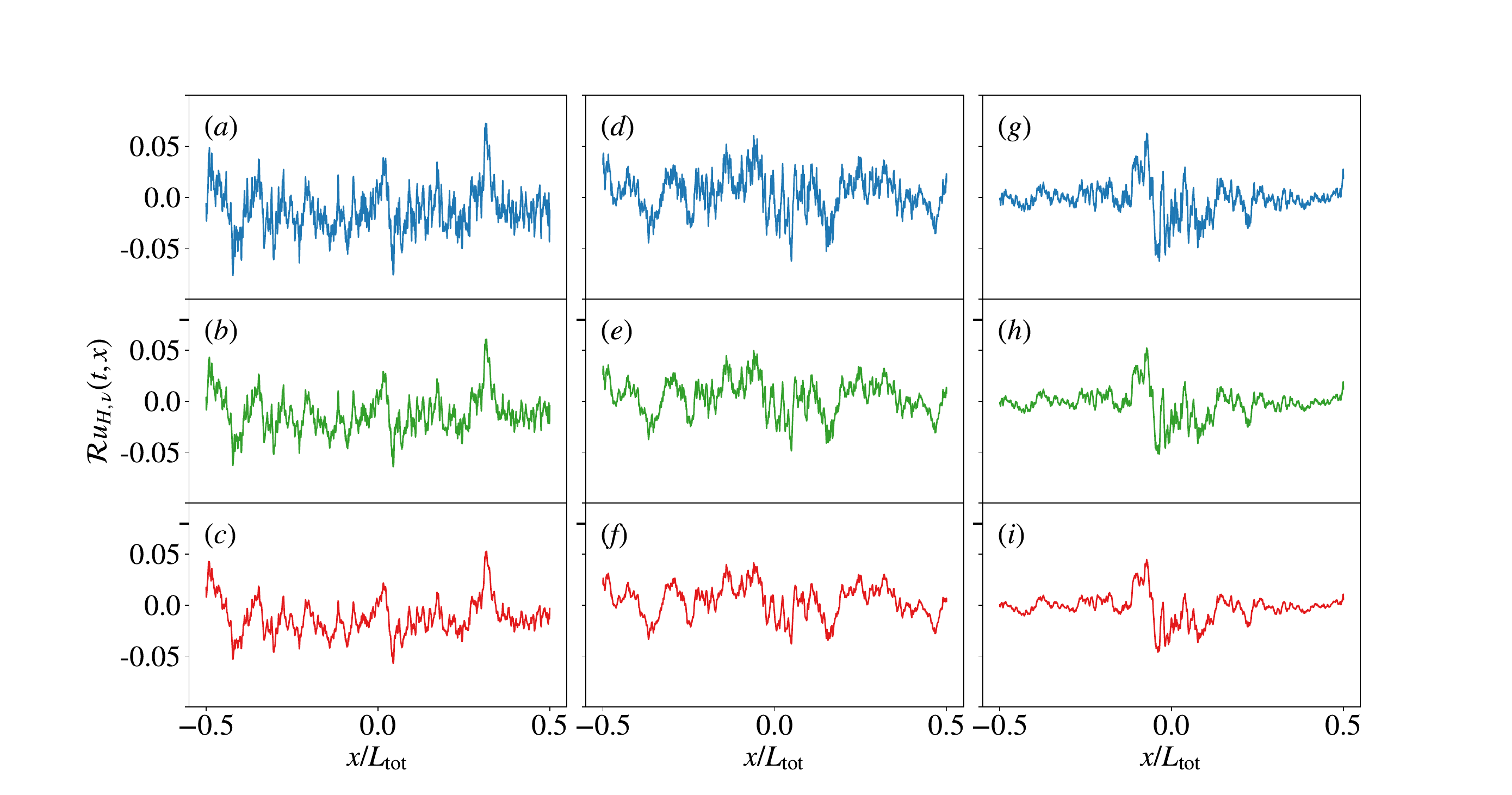}
\caption{Spatial profile of the real part of the numerical solution of the fractional equation (Eq.~\ref{eq:frac}) in the statistically stationary state, under an exponentially correlated in time forcing (Eq.~\ref{eq:ornstein-uhlenbeck}), for different correlation times. The parameters in the figures are $N=2^{12}$, $L=L_{\mathrm{tot}} / 8$, $c=1.0$ and $\nu = 2 \times 10^{-10}$. Three different values of the $H$ parameter  are shown: $H=1/3$ (blue), $H=1/2$ (green) and $H=2/3$ (red). Each figure corresponds to a different forcing correlation time, respectively, $T = 0.5$ (a,b,c), $T = 2.0$ (d,e,f) and $T = 5.0$ (g,h,i). For each value of $T$, the same space-time realization of the forcing is used for all values of $H$.}
\label{fig:exp-space-profile}
\end{figure}

In Fig.~\ref{fig:exp-space-profile}, the spatial profile of some of these numerical simulations is shown at a fixed time, in the stationary state. Analogously to Fig.~\ref{fig:delta-space-profile}, the same space-time realization of the forcing was applied to three different values of the H\"older parameter, and a different roughness can be observed for each curve. On the other hand, all of the profiles shown display the same value of viscosity, $\nu = 2 \times 10^{-10}$, with each figure corresponding to a different forcing correlation time, where in (a) it is the smallest, equal to $T = 0.5$, in (b) it is $T=2.0$ and in (c), $T=5.0$. The velocity fields in (a) look homogeneous and similar to those shown in Fig.~\ref{fig:delta-space-profile}(b), while in Fig.~\ref{fig:exp-space-profile}(b), as the correlation time grows, a tendency of accumulation at the center of the domain can be noticed, an effect that is confirmed in (c), where the nonhomogeneity of the system is clear and the correlation time is the largest. 

To understand these observations, let us consider the spatio-temporal correlation of the solution $u_{H,0}(t,x)$. After some calculations, we obtain
\beq\label{eq:STCorrExpForce}
 \mathbb{E}\left[u_{H,0}(t+\tau,x+\ell) u^*_{H,0}(t,x) \right]=\frac{1}{2T}\int_{s_1=0}^{t+\tau}\int_{s_2=0}^{t}e^{2i\pi c[s_1\ell+x(s_1-s_2)]}\mathcal D_\ell(s_1,s_2)e^{-\frac{|\tau+s_2-s_1|}{T}}ds_1ds_2,
\eeq
where enters the symmetric quantity
\beq \begin{split} \label{eq:ExpressDs1s2}
\mathcal D_\ell(s_1,s_2)&=\int_{k\in\mathbb R}e^{2i\pi k\ell}|k+cs_1|^{-H-1/2}_{1/L}|k+cs_2|^{-H-1/2}_{1/L}|k|^{2H+1}_{1/L} \widehat{C}_f(k)dk\\
&=\mathcal D_\ell(s_2,s_1).
\end{split} \eeq

Concerning the variance at large times, consider the expression provided in Eq. \eqref{eq:STCorrExpForce} for $\ell=0$ and $\tau=0$. Using the symmetry of the function $\mathcal D_{\ell}$, we get
\beq\label{eq:VarExpForce}
 \lim_{t \to \infty} \mathbb{E}\left|u_{H,0}(t,x) \right|^2=\frac{1}{2T}\int_{(s_1,s_2)\in (\mathbb R^+)^2}\cos(2\pi cx(s_1-s_2))\mathcal D_0(s_1,s_2)e^{-\frac{|s_2-s_1|}{T}} ds_1 ds_2,
\eeq
which depends explicitly on the position $x$ and is bounded by its value at the origin $x=0$. First, remark again that, for a frozen force, obtained by replacing $e^{-|s_1-s_2|/T}/2T \rightarrow 1$, this expression diverges at $x=0$ and $H<1/2$. Going back to the correlated in time forcing (Eq. \ref{eq:VarExpForce}), we can see its finiteness for $H>0$ while noticing that, along the diagonal $s_1=s_2$, using the expression of $\mathcal D_\ell$ (Eq. \ref{eq:ExpressDs1s2}) and the rapid decay of  $\widehat{C}_f(k)$, all integrals converge at large arguments. This explains the observations made in Fig. \ref{fig:exp-space-profile}.

\begin{figure}[t]
\centering
\includegraphics[width=\textwidth]{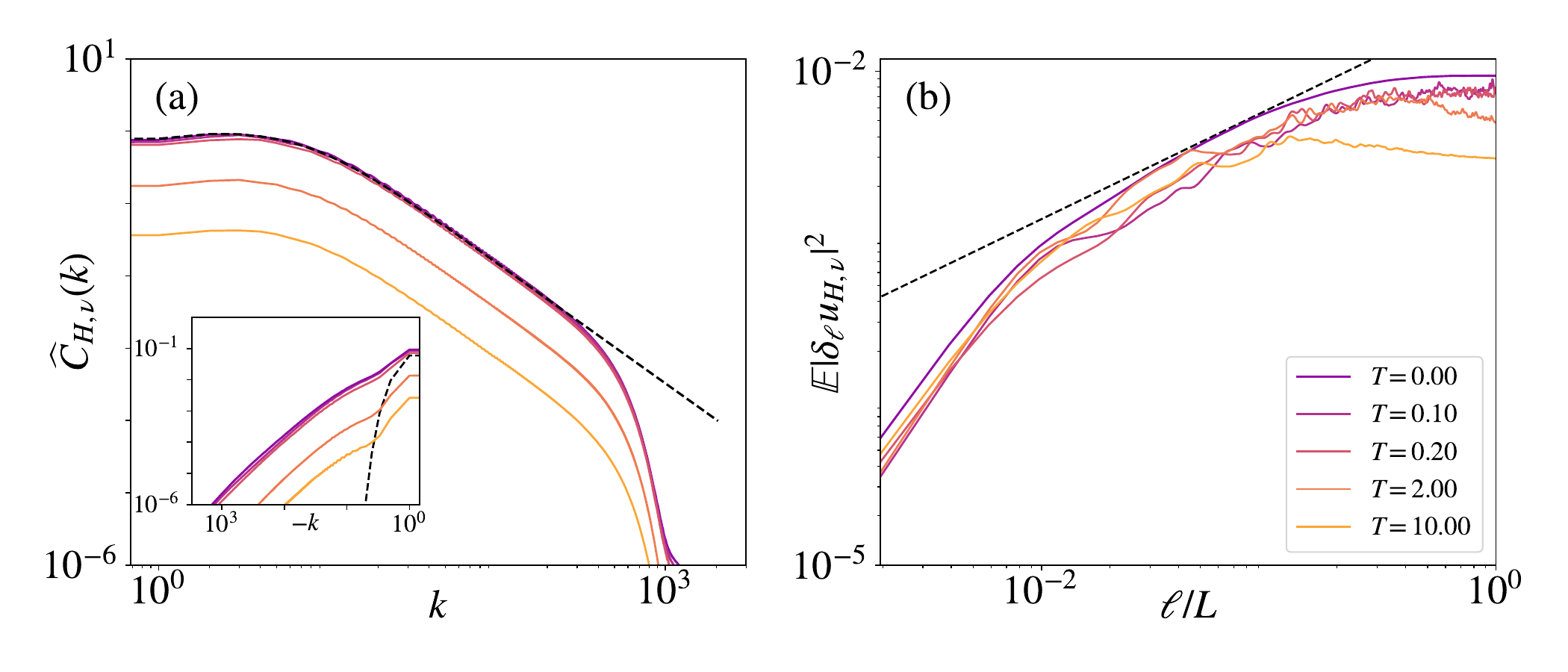}
\caption{Estimation of the spatial statistics of the fractional equation for different correlation times of the forcing, in comparison to a solution with delta-correlated forcing (in purple, indicated by $T=0$ in the legend). The parameters $H=1/3$ and $L=L_{\mathrm{tot}}/8$ are used and all correlation times are indicated in the legend. In (a), the power spectrum shown is estimated from the periodogram, as $\mathbb{E}|\widehat{u}_{H,\nu}|^2/N$, and the black dashed line corresponds to the analytical expression of the power spectrum in the inviscid and white-in-time case (Eq.~\ref{eq:delta-exact-spec}). The inset shows the corresponding negative frequencies of the power spectrum, with the same analytical expression in a black dashed line. In (b), the second order structure function, calculated with respect to the origin (see text) is shown, along with the analytical prediction for the inviscid delta-correlated equation (Eq.~\ref{eq:frac-sf2-space}).}
\label{fig:exp-space-stats}
\end{figure}

The spatial statistics of this velocity field is shown in Fig.~\ref{fig:exp-space-stats}, for different correlation times of the forcing. Similarly to Fig. \ref{fig:delta-space-stats}(a), we begin displaying in  Fig.~\ref{fig:exp-space-stats}(a) the periodogram. For a direct comparison, we reproduce the white-in-time forcing case, indicated by $T=0.0$ and plotted with a dark purple curve. For small values of the correlation time, $T=0.1$ and $T=0.2$, the curves can be hardly distinguished from the white-in-time case, both of them also matching the analytical prediction of Eq.~\eqref{eq:delta-exact-spec}, in a black dashed line. As the correlation time grows, the general shape of the curve is preserved, with an inertial range displaying power-law decay with an exponent of $-(2H+1)$, but at significantly smaller values, due to the $1/T$ prefactor in the time correlation of the force. In the inset, the power spectrum at negative frequencies is shown for the same parameters, and the agreement with the original simulation is observed for small correlation times, while smaller values with the same shape are observed for larger correlation times.

As we have seen, in particular while considering the variance in presence of a forcing term which is correlated in time (Eq. \ref{eq:VarExpForce}), the solution is not statistically homogeneous. This is why we calculate the second order structure function in space of the nonhomogeneous velocity fields with respect to the origin. That is, all velocity increments at scale $\ell$ are defined as
\beq \delta_{\ell} u_{H,\nu} = u_{H,\nu}(t,\ell) - u_{H,\nu}(t,0) \ , \eeq
and averaged only over independent instants in time. Our results are displayed in Fig.~\ref{fig:exp-space-stats}(b).
Being calculated only with respect to the origin, the statistics available is much smaller and the numerical curves are not as well resolved. Nevertheless, a tendency to follow the power-law behavior with a $2H$ exponent is still observed, and even the variance of the system at the origin, as measured by the large separation results, has a similar value to that of the $T=0$ curve. For larger correlation times, this agreement is lost, since the horizontal asymptote is $\mathbb{E}|u_{H,\nu}(t,\ell)|^2 + \mathbb{E}|u_{H,\nu}(t,0)|^2$, which differs significantly from $2 \mathbb{E}|u_{H,\nu}(t,0)|^2$, as already shown in Eq.~\eqref{eq:VarExpForce}. An analytical expression for the increment variance can be obtained from Eq. \eqref{eq:STCorrExpForce}:
\beq \begin{split}
\mathbb{E}|\delta_{\ell} u_{H,0}|^2 &= \frac{1}{T} \int_{(s_1,s_2) \in (\mathbb{R}^+)^2} \left( \mathcal{D}_0(s_1,s_2) - \mathcal{R} e^{2 i \pi c s_1 \ell} \mathcal{D}_{\ell}(s_1,s_2) \right) e^{-|s_1-s_2|/T} ds_1 ds_2 \ .
\end{split} \eeq
This expression is positive and real, as expected,
and in the limit $T \to 0$, the delta-correlated result, Eq.~\eqref{eq:frac-sf2-space}, is recovered.

\begin{figure}[t]
\centering
\includegraphics[width=\textwidth]{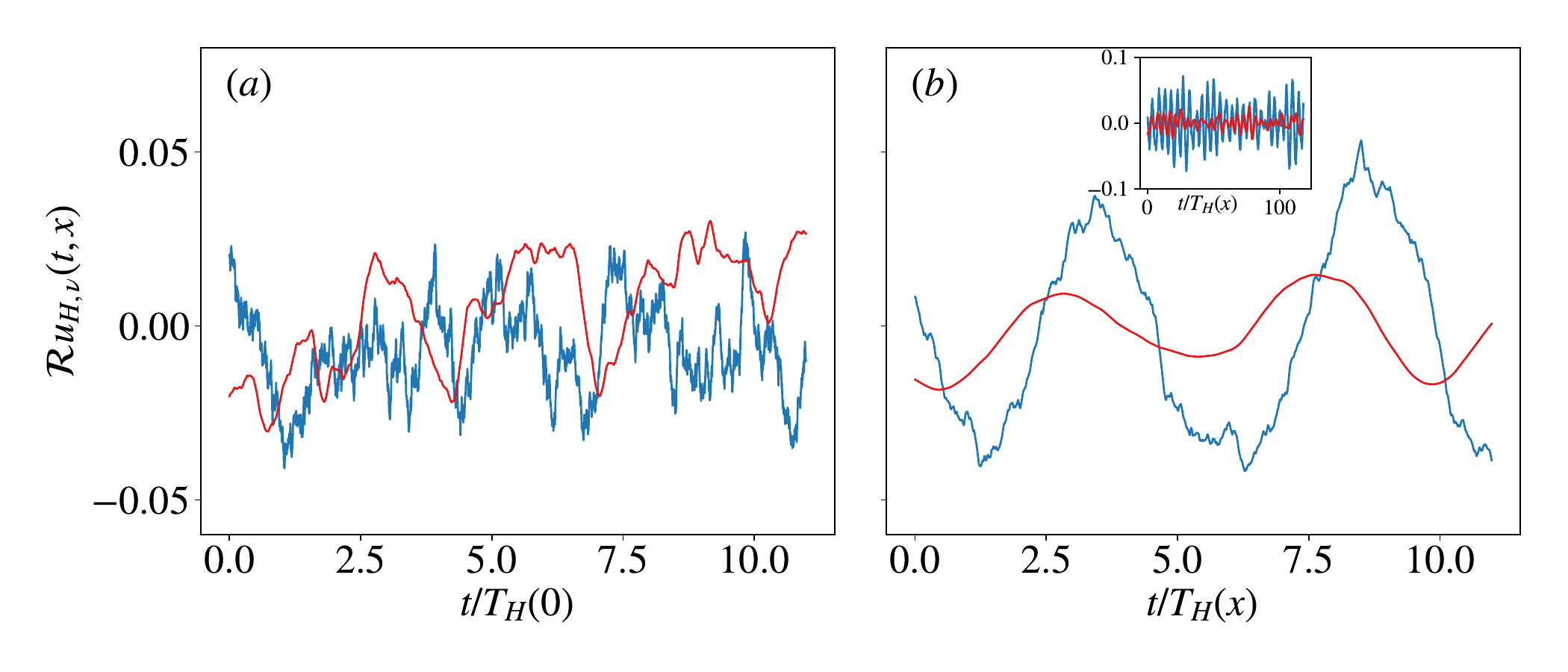}
\caption{Profile in time of the velocity field $u_{H,\nu}$ at the origin (a) and at $x=0.1 L_{\mathrm{tot}}$ (b), each rescaled by the local integral correlation time calculated for the white-in-time forcing case in Eq.~\eqref{eq:delta-int-time-off}. Each figure shows two values of the correlation time: $T=0.1$ (blue) and $T=10.0$ (red). The inset of (b) shows the same velocity profile as in (b), for a total time equal to the one of (a), in units of the simulation.}
\label{fig:exp-time-profile}
\end{figure}

The statistical properties in time of Eq.~\eqref{eq:frac} under correlated forcing in time are investigated in the next two figures. For these observables, we remind that their statistics under delta-correlated force was already nonhomogeneous, as shown in Eq.~\eqref{eq:delta-sf2-time-off} and discussed in Figs.~\ref{fig:delta-time-profile} and ~\ref{fig:delta-time-stats}. An important difference, though, is that the solution $u_{H,\nu}$ under correlated forcing is continuous and differentiable in time as well as in space, as can be seen in Fig.~\ref{fig:exp-time-profile}, where a profile in time is shown at different positions, the origin (a) and at $x = 0.1 L_{\mathrm{tot}}$ (b). In each figure, two values of the forcing correlation time are shown, $T=0.1$ (blue) and $T=10.0$ (red), and it can be noticed that, for the small correlation time, the curve looks rough exactly as the one in Fig.~\ref{fig:delta-time-profile}(a), its regularization in time happening only at very small scales, while the red curve, with a larger correlation time, is noticeably smoother. Outside the origin, as shown in Fig.~\ref{fig:exp-time-profile}(b), the horizontal axis has been rescaled by the corresponding integral time (Eq.~\ref{eq:delta-int-time-off}), and an equivalent interval as in (a) is shown. In the inset, a total time equal to the one in (a), in units of the simulation, is shown, and many multiples of the integral time can be seen. An important difference with the case of white-in-time forcing, though, is that the variance is almost homogeneous when the correlation time is small, while it decays fast for larger $T$, reason for which the red curve has a smaller amplitude than the blue curve, as can be clearly noticed in the inset.

\begin{figure}[t]
\centering
\includegraphics[width=\textwidth]{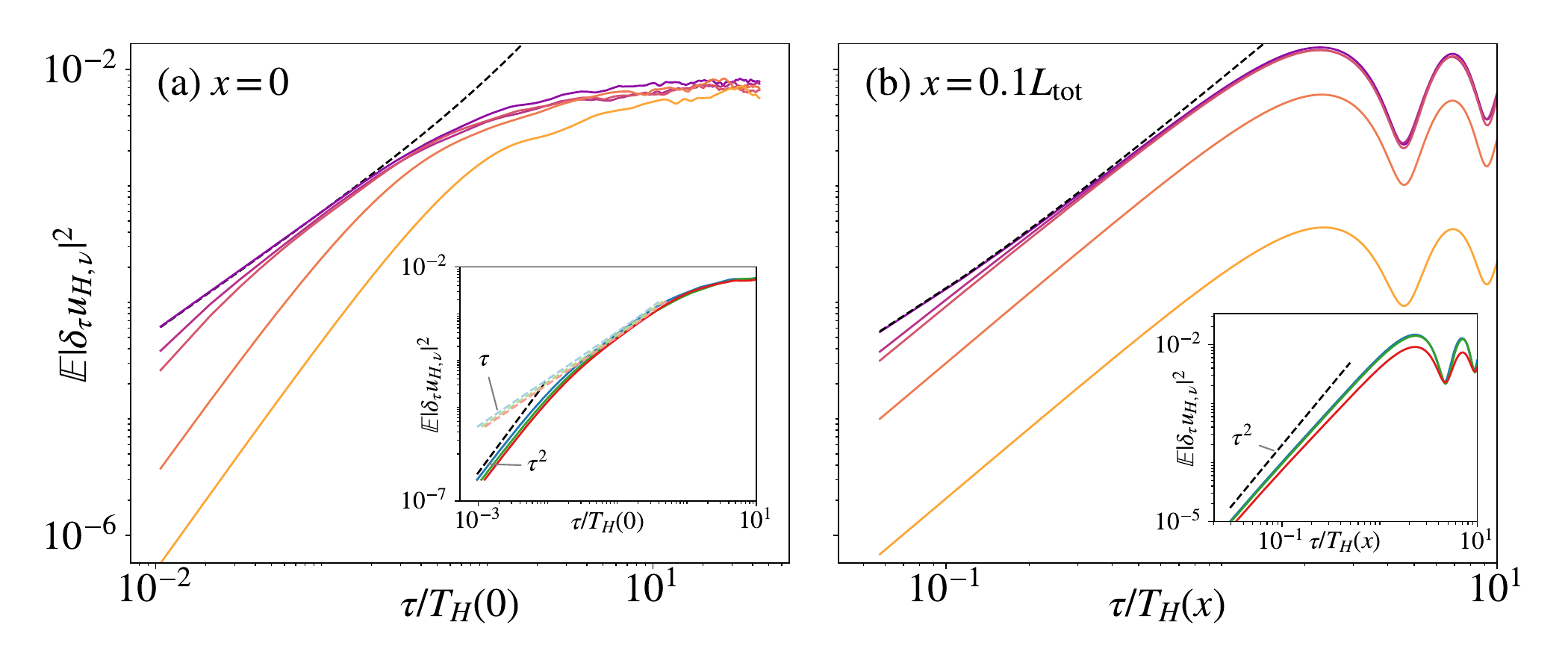}
\caption{Numerical estimation of the second order structure function in time of Eq.~\eqref{eq:frac} under correlated forcing in time. The colors follow the same pattern as in Fig.~\ref{fig:exp-space-stats}, where the purple curve represents a delta-correlated simulation and yellow represents the largest correlation time ($T=10.0$). In (a), this structure function is evaluated at the origin and in (b), at $x=0.1 L_{\mathrm{tot}}$. Each horizontal axis has been rescaled according to the local integral correlation time, as given by the estimate of Eq.~\eqref{eq:delta-int-time-off} and the black dashed curves are the same as in Fig.~\ref{fig:delta-time-stats}.
The inset of (a) shows the same structure function, at $x=0$, $\nu=2 \times 10^{-10}$ and $T=0.1$ for three different values of $H$, together with the analytical results for a delta-correlated forcing (dashed light-colored curves) and a curve proportional to $\tau^2$ (in black dashed). The same asymptotic $\tau^2$ behavior is shown in the inset of (b), which is calculated at $x=0.1 L_{\mathrm{tot}}$.}
\label{fig:exp-time-stats}
\end{figure}

The corresponding statistical properties in time are reported in Fig.~\ref{fig:exp-time-stats}, where we plot the temporal second order structure function for different correlation times of the forcing and at two different positions, the origin (a) and at $x = 0.1 L_{\mathrm{tot}}$ (b). In (a), the black dashed curve, corresponding to the small-time analytical result for the delta-correlated forcing (Eq.~\ref{eq:delta-sf2-time-off} with $x=0$) can be seen to match only the corresponding delta-correlated numerical simulation (dark purple) at very small time intervals. All other curves, with a finite correlation time, show a dissipative range in time, where this structure function is proportional to $\tau^2$ instead of $\tau$. This can be understood by looking at the derivatives of the correlation function in time in the stationary state. Consider the Taylor expansion of Eq.~\eqref{eq:STCorrExpForce} around $\tau$, with $\ell=0$. The first order contribution is proportional to
\beq \begin{split} \label{eq:corrf-time-taylor-first}
&\frac{d \mathbb{E}[ u_{H,0}(t+\tau,x) u_{H,0}^*(t,x) ]}{d \tau} = \frac{1}{2T} \int_{s_2=0}^{t} e^{2 \pi i c x (t+\tau-s_2)}\mathcal{D}_0(t+\tau,s_2) e^{-|t-s_2|/T} ds_2 \\&+
\frac{1}{2T^2} \int_{s_1=0}^{t+\tau} \int_{s_2=0}^{t} e^{2 \pi i c x (s_1-s_2)} \mathcal{D}_0(s_1,s_2) e^{-|\tau+s_2-s_1|/T} \frac{\tau+s_2-s_1}{|\tau+s_2-s_1|} ds_1 ds_2 \ .
\end{split} \eeq
The first term in this equation vanishes in the limit $t \to \infty$ and the second term is identically zero at $\tau=0$ because it is odd under the exchange of $s_1$ and $s_2$. Therefore, the contribution proportional to $\tau$ vanishes. Deriving Eq.~\eqref{eq:corrf-time-taylor-first} with respect to $\tau$, we obtain a single nonvanishing contribution in the stationary limit, equal to
\beq \label{eq:sf2-ou-prefac} \begin{split}
\lim_{t \to \infty} \frac{d^2 \mathbb{E}[ u_{H,0}(t+\tau,x) u_{H,0}^*(t,x) ]}{d \tau^2} \Bigg\lvert_{\tau=0} &= -\frac{1}{2T^3} \int_{(s_1,s_2) \in (\mathbb{R}^+)^2}  \cos(2\pi c x (s_1-s_2)) \\&\times \mathcal{D}_0(s_1,s_2) e^{-|s_1-s_2|/T} ds_1 ds_2 \ ,
\end{split} \eeq
which is proportional to the variance, and is thus finite for any $H > 0$. Moreover, this contribution is negative, since the autocorrelation function at $\tau$ must necessarily be smaller than the variance. This confirms the numerical observation in Fig.~\ref{fig:exp-time-stats} that the introduction of a finite time correlation in the forcing induces a smooth behavior of the velocity field at small time intervals.

This new dissipative range depends only on the correlation time and not on viscosity, but it can still be seen that, for small values of $T$, there is an inertial range where the linear behavior is recovered, in agreement with Eq.~\eqref{eq:delta-sf2-time-off} at $x=0$. Furthermore, we observe that the variance of all of these solutions is approximately the same, as can be seen in the large time separation end of the graph. Hence, at the origin the velocity field $u_{H,\nu}$ retains a variance close to that of the white-in-time scenario, as already observed in Fig.~\ref{fig:exp-space-stats} for small correlation times.

In Fig.~\ref{fig:exp-time-stats}(b), the second order structure function in time is shown at the position $x=0.1 L_{\mathrm{tot}}$. Again, it can be noticed that, for small values of the correlation time, the darker curves closely follow the behavior of the delta-correlated solution, but differ only at very small time increments, where the finitely correlated solutions show smooth behavior, proportional to $\tau^2$, while the white-in-time forcing solution resembles a Brownian motion in time, as given by Eq.~\eqref{eq:delta-sf2-time-off} at very small values of $\tau$. For larger correlation times, the same general behavior is observed, with a dissipative range at small time separations, oscillations with the same pattern as those of the delta-correlated forcing, but significantly smaller values of the variance, as given by the asymptotic values of these curves at large time. This is again a consequence of the loss of homogeneity of the solution $u_{H,\nu}$ as the correlation time of the force is increased.

The insets of Fig.~\ref{fig:exp-time-stats} evidence the fact that the power-law behavior of the temporal structure function only depends on the forcing correlation, and not on $H$. The light-colored dashed curves in the inset of (a) follow Eq.~\eqref{eq:delta-sf2-time-off}, for a delta-correlated in time forcing, and show a scaling proportional to $\tau$, whereas the numerical curves, with a finite correlation time of $T=0.1$, show a scaling proportional to $\tau^2$ at small time separations. This can be verified by comparison against an exact quadratic scaling represented in a black dashed line.
The inset of (b), calculated at $x=0.1 L_{\mathrm{tot}}$, also shows a quadratic scaling at small time intervals. As in the delta-correlated case, the prefactors show a slight $H$ dependence, but the scaling depends only on the temporal correlation of the forcing.

\section{Concluding remarks}

We have investigated the statistical properties in space and in time of a linear dynamical model of the turbulent energy cascade, introduced in Ref. \cite{apolinario2021}. This is a linear stochastic partial differential equation with a smooth large-scale forcing in space and general correlation in time. The spatial statistics under a white-in-time forcing were first investigated in \cite{apolinario2021}, where it is observed that the solution, at fixed times, is homogeneous and resembles a fractional Gaussian field with H\"older parameter $H$, a stationary state which is reached even in the absence of viscosity. Hence, with the specific choice of $H=1/3$, the spatial profiles of the solution offer a representation of the K41 framework of turbulence as the result of a stochastic PDE.
Nevertheless, the statistics in time under delta-correlated forcing, first investigated in the present work (Sec.~\ref{sec:fgf}), are not homogeneous, and are not influenced by the regularity in space, displaying the same local regularity in time as a Brownian process. At the origin of the system, the correlation time of the solution is larger, but it rapidly decays as one moves away from this point.  One can contrast this picture with turbulent velocity fields in the Eulerian setting, in which the same local roughness in time and in space is observed \cite{chevillard2005,gorbunova2021}.

The effects of a finitely correlated in time forcing were investigated in Sec.~\ref{sec:corrf}. For very small correlation times $T$, many of the previous results for a delta-correlated in time force are reproduced, but as $T$ is increased, the solution becomes clearly nonhomogeneous, as can be seen in its spatial profiles (Fig.~\ref{fig:exp-space-profile}). The local roughness of $H$ is preserved for small spatial scales, but these nonhomogeneous velocity fields are no longer a good portrait of turbulent flows. Since the forcing is now smooth in space and time, the solution is smooth in time as well, as shown by its second order structure functions, which display a dissipative range (Figs.~\ref{fig:exp-space-stats} and ~\ref{fig:exp-time-stats}). This dissipative regime in time is again not influenced by viscous dissipation or by the H\"older exponent of the solution.

This simple cascade model allows for the construction of fractional Gaussian fields in a dynamical setting, starting from a smooth force in space. In \cite{apolinario2021}, a nonlinearity which induces non-Gaussian statistics on this velocity field was investigated, but only for a delta-correlated in time forcing. Therefore, the combined effects of nonlinearity and correlations in time of the forcing in this model remain to be understood.
 In particular, we notice the contrast between the model and hydrodynamic turbulence, in which correlations in time of the stirring do not crucially change statistical properties of the solution.

\section*{Acknowledgments}
We thank Jules Guioth and Malo Tarpin for enriching discussions. The authors are partially supported by the Simons Foundation Award ID 651475.

\section*{Conflict of interest}

The authors declare no conflict of interest.

\bibliography{bibliography}

\end{document}